\documentclass[english,letterpaper]{revtex4}
\usepackage[T1]{fontenc}
\usepackage[latin1]{inputenc}
\usepackage{float}
\usepackage{graphicx}

\makeatletter

\providecommand{\tabularnewline}{\\}

\usepackage{float}

\makeatletter

\usepackage{float}

\makeatletter

\usepackage{geometry}

\usepackage{float}



\makeatother

\makeatother

\usepackage{babel}
\makeatother
\begin{document}

\title{A relativistic unitary coupled-cluster study of electric quadrupole
moment and magnetic dipole hyperfine constants of $\,^{199}Hg^{+}$}

\author{Chiranjib Sur}

\affiliation{Department of Astronomy, The Ohio State University, Columbus, Ohio,
43210 USA}

\email{csur@astronomy.ohio-state.edu}

\homepage{http://www.astronomy.ohio-state.edu/~csur}

\author{Rajat K Chaudhuri}

\affiliation{Indian Institute of Astrophysics, Koramangala, Block II, Bangalore,
560 034 INDIA}

\date{Journal Ref : Phys. Rev. A : 76, 032503 (2007)}

\begin{abstract}
Searching for an accurate optical clock which can serve as a better
time standard than the present day atomic clock is highly demanding
from several areas of science and technology. Several attempts have
been made to built more accurate clocks with different ion species.
In this article we discuss the electric quadrupole and hyperfine shifts
in the $5d^{9}6s^{2}\,^{2}D_{5/2}\left(F=0,m_{F}=0\right)\leftrightarrow5d^{10}6s\,^{2}S_{1/2}\left(F=2,m_{F}=0\right)$
clock transition in $\,^{199}Hg^{+}$, one of the most promising candidates
for next generation optical clocks. We have applied Fock-space unitary
coupled-cluster (FSUCC) theory to study the electric quadrupole moment
of the $5d^{9}6s^{2}\,^{2}D_{5/2}$ state and magnetic dipole hyperfine
constants of $5d^{9}6s^{2}\,^{2}D_{3/2,5/2}$ and $5d^{10}6s^{1}\,^{2}S_{1/2}$
states respectively of $\,^{199}Hg^{+}$. We have also compared our
results with available data. To the best of our knowledge, this is
the first time a variant of coupled-cluster (CC) theories has been
applied to study these kinds of properties of $Hg^{+}$and is the
most accurate estimate of these quantities to date.

~

\textbf{PACS number(s).} : 31.15.Ar, 31.15.Dv, 32.30.Jc, 31.25.Jf,
32.10.Fn 
\end{abstract}
\maketitle

\section{\label{intro}Introduction}

\begin{figure}[H]

\begin{centering}
\includegraphics[scale=0.6]{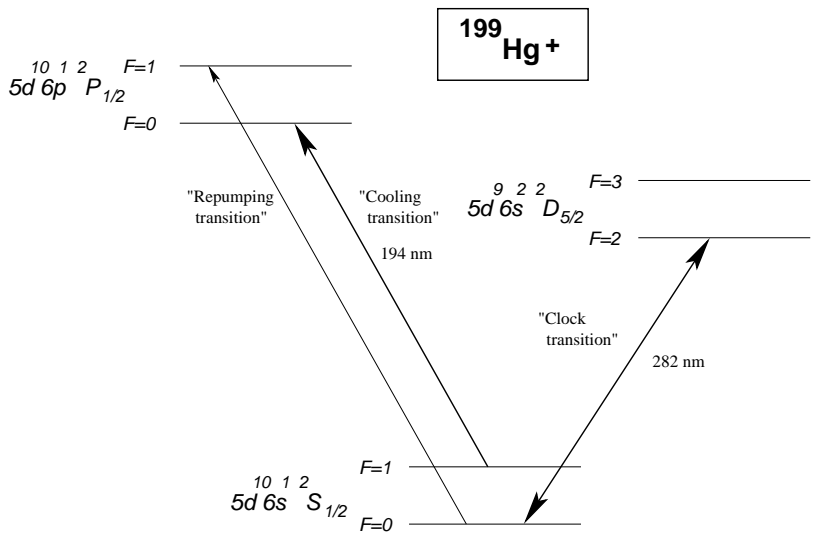} 
\par\end{centering}

\caption{\label{en-levels}Partial energy level diagram (including hyperfine
levels) of the $\,^{199}Hg^{+}$ion. The `clock-transition' ($5d^{9}6s^{2}\,^{2}D_{5/2}\left(F=0,m_{F}=0\right)\leftrightarrow5d^{10}6s\,^{2}S_{1/2}\left(F=2,m_{F}=0\right)$)
which is of interest, is a forbidden electric quadrupole ($E2$) type
at 282 $nm$.}
\end{figure}

The frequencies at which atoms emit or absorb electro-magnetic radiation
during a transition can be used for defining the basic unit of time.
The transitions that are extremely stable, accurately measurable and
reproducible can serve as excellent frequency standards. Present frequency
standard is based on the transition between the two hyperfine levels
of a cesium atom ($[Xe]6s\left(\,^{2}S_{1/2},F=3,m_{F}=0\right)\leftrightarrow[Xe]6s\left(\,^{2}S_{1/2},F=4,m_{F}=0\right)$)
with an accuracy of 1 part in $10^{15}$. However, there is an ongoing
search for even more accurate clocks in the optical regime. Recent
day progress in technologies makes it feasible to built more accurate
clocks with higher precision. Moreover this kind of study not only
provides the foundation for a wide range of experiments and precision
measurements but also can be used for stringent tests of our fundamental
concepts and theories. Some recent studies of frequency standards
have yielded sensitive probes of possible temporal variation of the
fundamental constants \cite{marion-03,bize-03,peik-04}. Atomic frequency
standards based on a single trapped ion has been established to provide
more stability and accuracy than those of present-day time standards
\cite{csur-tc2005}. Among all the ionic candidates for frequency
standards, $\,^{199}Hg^{+}$ \cite{nist-fs} is believed to be the
most reliable one. Recent progress on $\,^{199}Hg^{+}$ frequency
standards \cite{nist-fs,progress-03,progress-06} have revealed the
feasibility of achieving an accuracy of $1$ part in $10^{18}$ as
compared to $1$ part in $10^{15}$ which is the present day standard.
In particular, the 282 $nm$ transition ($5d^{9}6s^{2}\,^{2}D_{5/2}\left(F=0,m_{F}=0\right)\leftrightarrow5d^{10}6s\,^{2}S_{1/2}\left(F=2,m_{F}=0\right)$)
is of interest in $\,^{199}Hg^{+}$. A schematic diagram of the energy
levels and the clock transition is given in Fig. (\ref{en-levels}).
The electronic structure of $\,^{199}Hg^{+}$ reveals that the excited
(metastable) states which are interesting from the point of view of
frequency standards, involve open $d$-shells and are very difficult
to evaluate using any theoretical methods. Therefore, any kind of
property calculation of $\,^{199}Hg^{+}$ involving an open $d$-
shell is very complex and challenging.

To measure the transition frequency accurately one needs to determine
the corresponding states (energy levels) with a high precision. When
an atom interacts with an external field, the standard frequency may
be shifted from the resonant frequency. The quality of the frequency
standard depends upon the accurate and precise measurement of this
shift. To minimize or maintain any shift of the clock frequency, the
interaction of the atom with it's surroundings must be controlled.
Hence, it is important to have a good knowledge of these shifts so
as to minimize them while setting up the frequency standard. Some
of these shifts are the linear and quadratic Zeeman shift, second-order
Stark shift, hyperfine shift and electric quadrupole shift. The largest
source of uncertainty in frequency shift arises from the electric
quadrupole shift and the hyperfine shift of the clock transition.
Departure of the spherical symmetries in the $D$ states of $\,^{199}Hg^{+}$
gives rise to an electric quadrupole moment and in the presence of
external electric field gradient the atomic electric quadrupole moment
will cause a shift in the energy levels of the $D$ states. On the
other hand, the non-zero nuclear spin of $\,^{199}Hg^{+}$ produces
nuclear multipole moments which interacts with the electron moments
at the site of the nucleus which is caused by the nuclear spin. This
interaction will lead to a hyperfine effect and the corresponding
shift in the upper and lower levels are known as hyperfine shifts.

In this article we have used the relativistic Fock-space unitary coupled
cluster (FSUCC) method, one of the most accurate theories to describe
the electron correlation effects in many-electron atoms, to calculate
electric quadrupole moment and hyperfine constants of $\,^{199}Hg^{+}$.
The excited states which are of interests in $\,^{199}Hg^{+}$involve
open $d$- shells which make the calculation very complex and challenging.
Unlike ordinary Fock-space coupled-cluster (FSCC) theory, FSUCC is
based on unitary groups and contains much more physical effects in
the same level of approximation. Although the electric quadrupole
moment (EQM) has been determined by the experimentalists and the theoreticians,
to the best of our knowledge, this is the first time a theory of this
kind is being applied to study the properties of a complicated system
like \textbf{$\,^{199}Hg^{+}$.} The precise determination of the
hyperfine constants using FSUCC theory can be considered as benchmarking
of the determination of atomic states. Therefore, the accuracy obtained
for the EQMusing this approach can help us to determine the uncertainty
(which is $\sim3.5\%$for the experiment) of the same which will be
very useful in the frequency standard studies. 

The structure of the paper is as follows : Section \ref{intro} gives
a brief introduction about the importance of using $\,^{199}Hg^{+}$
in frequency standards. It also introduces the importance of applying
FSUCC theory in this problem. This is followed by Sec. \ref{theory}
which deals with a short theoretical description of FSUCC theory,
electric quadrupole moment and hyperfine structure. We present the
results and the relevant discussions in the next two Sec. \ref{results}.
Finally, in Sec. \ref{concl} we conclude and highlight the important
findings of our work.

\section{\label{theory}Theoretical details}

In this section we describe the theoretical formulation of our work.
Relativistic Fock-space unitary coupled-cluster (FSUCC) theory is
a variant of well known Fock-space coupled-cluster (FSCC) theory in
the relativistic regime which is based on an unitary ansatz. More
details about UCC theory for atoms can be found in Ref. \cite{csur-ucc}
which is referred as I in the following sections. Therefore, we have
just outlined the FSUCC theory here. We obtain the open shell $5d^{9}6s^{2}\,^{2}D_{5/2}$
and $5d^{9}6s^{2}\,^{2}D_{3/2}$ states of $\,^{199}Hg^{+}$ using
core ionization technique. In one of our earlier papers, we have outlined
the treatment of open-shell coupled-cluster core ionization potential
(OSCC-IP) \cite{csur-coreip} which is referred as II in this manuscript.
For determining the $5d^{10}6s\,^{2}S_{1/2}$ state of $\,^{199}Hg^{+}$
we have used the open-shell coupled-cluster with electron attachment
(OSCC-EA) method \cite{mukherjee-oscc}. For more details we refer
to the original article by Lindgren and Mukherjee \cite{mukherjee-oscc}.

\subsection{\label{fsucc}Fock-space unitary coupled cluster theory}

\noindent Relativistic coupled-cluster (RCC) theory is based on the
no-virtual-pair approximation along with the appropriate modification
of the orbital form and potential terms \cite{eliav}. One begins
with the Dirac-Coulomb Hamiltonian ($H$) for an $N$-electron atom
which is expressed as

\begin{equation}
H=\sum_{i=1}^{N}\left[c\vec{\alpha_{i}}\cdot\vec{p}_{i}+\beta mc^{2}+V_{\mathrm{Nuc}}(r_{i})\right]+\sum_{i<j}^{N}\frac{e^{2}}{r_{ij}}\,.\label{dc}\end{equation}
 with all the standard notations often used. In FSCC/FSUCC method,
the self-consistent field solution of the Hartree-Fock (Dirac-Fock
in relativistic regime) for the $N$-electron closed shell ground
state $\Phi$ is chosen as the vacuum (for labeling purpose only)
to define holes and particles with respect to $\Phi$. The multi-reference
aspect is then introduced by subdividing the hole and particle orbitals
into active and inactive categories, where different occupations of
the active orbitals will define a multi-reference \emph{model} space
for our problem. We call a model space to be \emph{complete} if it
has all possible electron occupancies in the active orbitals, otherwise
incomplete. The classification of orbitals into active and inactive
groups is, \emph{in principle}, arbitrary and is at our disposal.
However, for the sake of computational convenience, we treat only
a few hole and particle orbitals as active, namely those are close
to the Fermi level.

We designate by $\Psi_{i}^{0(k,l)}$ a model space of $k$-hole and
$l$-particle determinants, where in the present instance ($\mathrm{Hg}-e\longrightarrow\mathrm{Hg^{+}}$),
$l=0$ and $k$ ranges from 0 to 1. Generally, any second quantized
operator has $k$-hole and $l$-particle annihilation operators for
the active holes and particles. For convenience, we indicate the {}``hole-particle
valence rank'' of an operator by a superscript ($k,l$) on the operator.
Thus, according to our notation, an operator $A^{(k,l)}$ will have
exactly $k$-hole and $l$-particle annihilation operators.

We now describe the type of ansatz used to derive the FSUCC equations
for direct energy difference calculations in one-electron detachment
processes. The Hartree-Fock/Dirac-Fock function $\Phi$ is denoted
by $\Psi^{(0,0)}$ and the inactive hole and particle orbitals (defined
with respect to $\Phi$) are labeled by the indices $a,b,c,\cdots$
and $p,q,r,\cdots$, respectively. The corresponding active holes
and particles are labeled by the indices $\alpha,\beta,\gamma\cdots$
and $u,v,w\cdots$, respectively. Note that there will be \emph{no
active} particles for electron detachment processes. The cluster operator
correlating the N-electron ground/reference state is denoted in our
notation by $S^{(0,0)}$ which can be split into various $n$-body
components depending upon the various hole-particle excitation ranks.
The cluster operator $S^{(0,0)}$ upto two-body can be written in
second quantized notation as, \begin{equation}
S^{(0,0)}=S_{1}^{(0,0)}+S_{2}^{(0,0)}+\cdots=\sum_{p,a}\langle p|s_{1}^{(0,0)}|a\rangle{\{ a_{p}^{\dagger}a_{a}}\}+\frac{1}{4}\sum_{a,b,p,q}\langle pq|s_{2}^{(0,0)}|ab\rangle{\{ a_{p}^{\dagger}a_{q}^{\dagger}a_{b}a_{a}}\}+\cdots\label{eq1}\end{equation}
 where $a^{\dagger}$ ($a$) denotes creation (annihilation) operator
with respect to $\Phi_{\mathrm{HF/DF}}$ and ${\{\cdots}\}$ denotes
\emph{normal} ordering. It should be noted that $S^{(0,0)}$ cannot
destroy any holes or particles; acting on $\Phi$, it can only create
them.

For $(N-1)$ electron states the model space consists of one active
hole and zero active particle \textbf{($k=1,l=0$)} and hence according
to our notation the valence sector for $(N-1)$ electron states can
be written as (1,0) sector. We introduce an wave operator $\Omega$
which generates all valid excitation from the model space function
for $(N-1)$ electron states. The wave operator $\Omega$ for the
(1,0) valence problem is given by \begin{equation}
\Omega={\{\exp(S^{(0,0)}+S^{(1,0)})}\}.\label{eq2}\end{equation}

\begin{figure}
\begin{centering}
\includegraphics{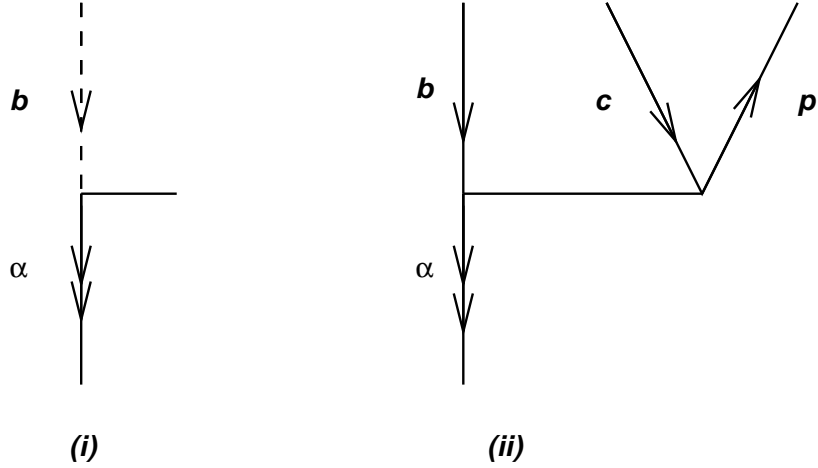} 
\par\end{centering}

\caption{\label{s-diag}Diagrammatic representation of the open-shell cluster
operators in the single and double excitation (SD) approximation.
Fig (i) and (ii) corresponds to single and double excitations amplitudes
respectively as given in Eq. (\ref{eq3}). The lines represented by
double arrow stand for occupied valence orbital and the downward single
arrows are occupied orbitals (including the valence) ; the line with
an upward arrow corresponds to particle. The dashed lines represent
inactive orbitals. }
\end{figure}

In this case the additional cluster operator $S^{(1,0)}$ must be
able to destroy the active particle present in the (1,0) valence space.
Like $S^{(0,0)}$, the cluster operator $S^{(1,0)}$ can also be split
into various $n$-body components depending upon hole-particle excitation
ranks. The one- and two-body $S^{(1,0)}$ (depicted in Fig.(\ref{s-diag}))
can be written in the second quantized notation as \begin{equation}
S^{(1,0)}=S_{1}^{(1,0)}+S_{2}^{(1,0)}+\cdots=\sum_{\alpha\ne b}\langle\alpha|s_{1}^{(1,0)}|b\rangle{\{ a_{\alpha}^{\dagger}a_{b}}\}+\frac{1}{2}\sum_{p,b,c}\langle p\alpha|s_{2}^{(1,0)}|cb\rangle{\{ a_{p}^{\dagger}a_{\alpha}^{\dagger}a_{b}a_{c}}\}+\cdots\label{eq3}\end{equation}
 where $\alpha$ denotes the active particle which is destroyed.

In general, for a $(k,l)$ valence problem, the cluster operator must
be able to destroy any subset of $k$- active holes and $l$- active
particles. Hence, the wave operator $\Omega$ for $(k,l)$ valence
sector may be written as \begin{equation}
\Omega={\{\exp({\tilde{S}}^{(k,l)})}\}.\label{eq6}\end{equation}
 where \begin{equation}
{\tilde{S}}^{(k,l)}={\sum_{m=0}^{k}}\hspace{0.1in}{\sum_{n=0}^{l}}S^{(m,n)}\label{eq7}\end{equation}

We define the exact wave function $\Psi_{i}^{(k,l)}$ for ($k,l$)
valence sector as \begin{equation}
\Psi_{i}^{(k,l)}=\Omega\Psi_{i}^{0(k,l)}\label{eq10}\end{equation}
 where \begin{equation}
\Psi_{i}^{0(k,l)}=\sum_{i}C_{i}^{(k,l)}\Phi_{i}^{(k,l)}.\label{eq11}\end{equation}
 The functions $\Phi_{i}^{(k,l)}$ in Eq.(\ref{eq11}) are the determinants
included in the model space $\Psi_{i}^{0(k,l)}$ and $C^{(k,l)}$
are the corresponding coefficients. Substituting the above form of
the wave-function (given in Eqs. (\ref{eq10}) and (\ref{eq11}))
in the Schr{ö}dinger equation for a manifold of states $H|\Psi_{i}^{(k,l)}\rangle=E_{i}|\Psi_{i}^{(k,l)}\rangle$,
we get \begin{equation}
H\Omega{\left(\sum_{i}C_{i}|\Phi_{i}^{(k,l)}\rangle\right)}=E_{i}\Omega{\left(\sum_{i}C_{i}|\Phi_{i}^{(k,l)}\rangle\right)},\label{eq9}\end{equation}
 where $E_{i}$ is the $i$-th state energy.

Following Lindgren \cite{Lindgren}, Mukherjee \cite{Mukherjee},
Lindgren and Mukherjee \cite{LindMukh}, Sinha \emph{et al.} \cite{Sinha}
and Pal \emph{et al.} \cite{Pal}, the Fock-space Bloch equation for
the FSCC may be written as \begin{equation}
H\Omega P^{(k,l)}=P^{(k,l)}H_{\mathrm{eff}}^{(k,l)}\Omega P^{(k,l)}\hspace{0.2in}\forall(k,l)\label{eq12}\end{equation}
 where \begin{equation}
H_{\mathrm{eff}}^{(k,l)}=P^{(k,l)}\Omega^{-1}H\Omega P^{(k,l)}\label{eq13}\end{equation}
 and $P^{(k,l)}$ is the model space projection operator for the ($k,l$)
valence sector (defined by $\sum_{i}C_{i}^{(k,l)}\Phi_{i}^{(k,l)}$).
For complete model space, the model space projector $P^{(k,l)}$ satisfies
the \emph{intermediate} normalization condition \begin{equation}
P^{(k,l)}\Omega P^{(k,l)}=P^{(k,l)}.\label{eq14}\end{equation}

At this juncture, we single out the cluster amplitudes $S^{(0,0)}$
and call them $T$. The rest of the cluster amplitudes will henceforth
be called $S$. The normal ordered definition of $\Omega$ enables
us to rewrite Eq.(\ref{eq7}) as

\begin{equation}
\Omega=\exp(T)\{\exp(S)\}=\exp(T)\Omega_{v}\label{eq15}\end{equation}
 where $\Omega_{v}$ represents the wave-operator for the valence
sector.

To formulate the theory for direct energy differences, we pre-multiply
Eq.(\ref{eq12}) by $\exp(-T)$ and get \begin{equation}
\overline{H}\Omega_{v}P^{(k,l)}=\Omega_{v}P^{(k,l)}H_{\mathrm{eff}}^{(k,l)}P^{(k,l)}\ ,\,\,\,\,\,\forall(k,l)\ne(0,0)\label{eq16}\end{equation}
 where $\overline{H}=\exp(-T){H}\exp(T)$. Since $\overline{H}$ can
be partitioned into a connected operator $\widetilde{H}$ and $E_{\mathrm{ref/gr}}$
($N$-electron closed-shell reference or ground state energy), we
likewise define $\widetilde{H}_{\mathrm{eff}}$ as \begin{equation}
\widetilde{H}_{\mathrm{eff}}^{(k,l)}=H_{\mathrm{eff}}^{(k,l)}-E_{\mathrm{gr}},\,\,\,\,\,\forall(k,l)\ne(0,0)\label{eq17}\end{equation}

Substituting Eq.(\ref{eq17}) in Eq.(\ref{eq16}) we obtain the Fock-space
Bloch equation for energy differences: \begin{equation}
\widetilde{H}\Omega_{v}P^{(k,l)}=\Omega_{v}P^{(k,l)}\widetilde{H}_{\mathrm{eff}}^{(k,l)}P^{(k,l)},\,\,\,\,\,\forall(k,l)\ne(0,0)\label{eq18}\end{equation}
 Eqs. (\ref{eq12}) and (\ref{eq18}) are solved by the Bloch projection
method for $k=l=0$ and $k=0,l\ne0$, respectively, involving the
left projection of the equations with $P^{(k,l)}$ and its orthogonal
complement $Q^{(k,l)}$ ($P^{(k,l)}+Q^{(k,l)}$=1) to obtain the effective
Hamiltonian and the cluster amplitudes, respectively. At this point,
we recall that the cluster amplitudes in FSCC are solved hierarchically
through the \emph{subsystem embedding condition} \cite{SEC,Haque}
which is equivalent to the \emph{valence universality} condition used
by Lindgren \cite{Lindgren} in his formulation. For example, in the
present application, we first solve the FSCC for $k=l=0$ to obtain
the cluster amplitudes $T$. The operator $\widetilde{H}$ and $\widetilde{H}_{\mathrm{eff}}^{(1,0)}$
are then constructed from this cluster amplitudes $T$ to solve Eq.
(\ref{eq18}) for $k=1$, $l=0$ to determine $S^{(1,0)}$ amplitudes.
$\widetilde{H}$ is then diagonalized within the model space to obtained
the desired eigenvalues and eigenvectors. The diagonalization is followed
from the eigenvalue equation

\begin{equation}
\widetilde{H}_{\mathrm{eff}}^{(1,0)}C^{(1,0)}=C^{(1,0)}E.\label{eq19}\end{equation}
 where \begin{equation}
\widetilde{H}_{\mathrm{eff}}^{(1,0)}=P^{(1,0)}[\widetilde{H}+\overbrace{\widetilde{H}S^{(1,0)}}]P^{(1,0)}.\label{eq20}\end{equation}
 The expression $\overbrace{AB}$ in Eq.(\ref{eq20}) indicates that
operators $A$ and $B$ are connected by common orbital(s).

In FSUCC, the wave operator $\Omega$ in Eq.(\ref{eq15}) is replaced
by \begin{equation}
\Omega=\exp(\sigma)\left\{ \exp\left(S\right)\right\} =\Omega_{c}\Omega_{v}\label{wave-2}\end{equation}
 with the cluster operator defined as $\sigma=T-T^{\dagger}$. In
comparison to the ordinary coupled cluster (CC) theory the wave operator
$\sigma$ in UCC contains de-excitation operator ($T^{\dagger}$)
as well and therefore UCC theory contains more higher order effects
than the conventional CC theory in the same level of approximation.
In this work we have used unitary coupled-cluster theory in the single
and double excitation approximation to treat the closed shell correlation
consistently. We refer to the article I for further details.

\subsection{\label{EQshift}Electric quadrupole shift}

The interaction of the atomic quadrupole moment with the external
electric-field gradient is similar to the interaction of a nuclear
quadrupole moment with the electric fields generated by the atomic
electrons inside the nucleus. In the presence of the electric field,
this interaction gives rise to an energy shift by coupling with the
gradient of the electric field. The quadrupole moment ${\bf \Theta}$
of an atomic state $|\Psi(\gamma JFM_{F})\rangle$ is defined as the
diagonal matrix element of the quadrupole operator with the maximum
value $M_{J}$ and is expressed as

\begin{equation}
{\bf \Theta}=\left\langle \Psi(\gamma JFM_{F})\right|\Theta_{zz}\left|\Psi(\gamma JFM_{F})\right\rangle \,.\label{theta}\end{equation}
 Here $\gamma$ specifies the electronic configuration of the atoms
which distinguishes the initial and final states; $J$ is the total
angular momentum of the atom and $F$ is the summation of nuclear
and atomic angular momentum with $M_{F}$ its projection. The electric
quadrupole operator in terms of the electronic coordinates is given
by

\begin{equation}
\Theta_{zz}=-\frac{e}{2}\sum_{j}\left(3z_{j}^{2}-r_{j}^{2}\right),\label{theta-zz}\end{equation}
 where the sum is over all the electrons and $z$ is the coordinate
of the $j$th electron. To calculate the quantity we express the quadrupole
operator in its single particle form as

\begin{equation}
\Theta_{m}^{(2)}=\sum_{m}q_{m}^{(2)}.\label{theta-single}\end{equation}
 The single particle reduced matrix element of the electric quadrupole
operator is expressed as \cite{grant-rad} \begin{equation}
\left\langle j_{f}\right\Vert q_{m}^{(2)}\left\Vert j_{i}\right\rangle =\left\langle j_{f}\right\Vert C_{m}^{(2)}\left\Vert j_{i}\right\rangle \int dr\, r^{2}\left(\mathcal{P}_{f}\mathcal{P}_{i}+\mathcal{Q}_{f}\mathcal{Q}_{i}\right).\label{quad-eq}\end{equation}
 In Eq.($\ref{quad-eq}$), the subscripts $f$ and $i$ correspond
to the final and initial states respectively; $\mathcal{P}$ and $\mathcal{Q}$
are the radial part of the large and small components of the single
particle Dirac-Fock wavefunctions respectively and $j_{i}$ is the
total angular momentum for the $i$th electron. The angular factor
is given in by \begin{eqnarray}
\left\langle j_{f}\right\Vert C_{m}^{(k)}\left\Vert j_{i}\right\rangle = &  & (-1)^{(j_{f}+1/2)}\sqrt{(2j_{f}+1)}\sqrt{(2j_{i}+1)}\nonumber \\
 &  & \times\left(\begin{array}{ccc}
j_{f} & 2 & j_{i}\\
-1/2 & 0 & 1/2\end{array}\right)\pi(l,k,l^{\prime})\label{ang}\end{eqnarray}
 where \[
\pi(l,k,l^{\prime})=\left\{ \begin{array}{c}
\begin{array}{cc}
1 & \mathrm{if}\: l+k+l^{\prime}\,\,\mathrm{even}\\
0 & \mathrm{otherwise~;}\end{array}\end{array}\right.\]
 $l$ and $k$ being the orbital angular momentum and the rank of
the interaction respectively.

Finally using the Wigner-Eckart theorem we define the electric quadrupole
moment in terms of the reduced matrix elements as \begin{equation}
\left\langle j_{f}\right|\Theta_{m}^{(2)}\left|j_{i}\right\rangle =(-1)^{j_{f}-m_{f}}\left(\begin{array}{ccc}
j_{f} & 2 & j_{i}\\
-m_{f} & 0 & m_{f}\end{array}\right)\left\langle j_{f}\right\Vert \Theta^{(2)}\left\Vert j_{i}\right\rangle \label{wig-eck}\end{equation}
 More details about the evaluation of the electric quadrupole moment
using RCC theory is described in one of our recent papers \cite{csur-prl}.
The electric quadrupole shift is evaluated using the relation \cite{dube-Q}

\begin{equation}
\left\langle \Psi(\gamma JFM_{F})\right|\Theta\left|\Psi(\gamma JFM_{F})\right\rangle =\frac{-2A\left[3M_{F}^{2}-F(F+1)\right]\left\langle \Psi(\gamma JF)\right\Vert \Theta^{(2)}\left\Vert \Psi(\gamma JF)\right\rangle }{\left[(2F+3)(2F+2)(2F+1)2F(2F-1)\right]^{1/2}}\times\mathcal{O}(\alpha,\beta)\label{EQshift}\end{equation}
 and

\begin{equation}
\mathcal{O}(\alpha,\beta)=\left[(3\cos^{2}\beta-1)-\epsilon(\cos^{2}\alpha-\sin^{2}\alpha)\right].\label{alpha-beta}\end{equation}
 Here $\alpha$ and $\beta$ are two of the three Euler angles that
take the principal-axis frame of the electric field gradient to the
quantization axis and $\epsilon$ is an asymmetry parameter of the
electric potential function \cite{dube-Q}.

\subsection{\label{mag-dip-shift}Magnetic dipole hyperfine shift}

The non-zero nuclear spin gives rise to nuclear multipole moments
which interact with the electric multipole moments generated by the
atomic electrons at the site of the nucleus and this interaction is
collectively known as hyperfine interaction \cite{lindgren-book}.
Theoretical determination of hyperfine constants is one of the most
stringent tests of accuracy of the atomic wave functions near the
nucleus. Also accurate predictions of hyperfine coupling constants
require a precise incorporation of relativistic and correlation effects.
Like the $\,^{201}Hg^{+}$ isotope, $\,^{199}Hg^{+}$ also has a non
zero nuclear spin ($I=\frac{1}{2}$) and the $m_{F}=0$ levels for
both the $^{2}S_{1/2}$ and $^{2}D_{5/2}$ states are independent
of the first order Zeeman shift.

The hyperfine interaction is given by

\begin{equation}
H_{hfs}=\sum_{k}M^{(k)}\cdot\mathcal{T}^{(k)},\label{HYP-eqn}\end{equation}
 where $M^{(k)}$ and $\mathcal{T}^{(k)}$ are spherical tensors of
rank $k$, which corresponds to nuclear and electronic parts of the
interaction respectively. The lowest $k=0$ order represents the interaction
of the electron with the spherical part of the nuclear charge distribution.

In the first order perturbation theory, the energy corresponding to
the hyperfine interaction of the fine structure state are the expectation
values of $H_{hfs}$ such that

\begin{equation}
\begin{array}{ccc}
W(J) & = & \left\langle IJFM_{F}\right|{\displaystyle \sum_{k}}M^{(k)}\cdot\mathcal{T}^{(k)}\left|IJFM_{F}\right\rangle \\
 & = & {\displaystyle \sum_{k}}(-1)^{I+J+F}\left\{ \begin{array}{ccc}
I & J & F\\
J & I & k\end{array}\right\} \left\langle I\right\Vert M^{(k)}\left\Vert I\right\rangle \left\langle J\right\Vert \mathcal{T}^{(k)}\left\Vert J\right\rangle \,.\end{array}\label{hyp-energy}\end{equation}
 Here $\mathbf{I}$ and $\mathbf{J}$ are the total angular angular
momentum for the nucleus and the electron state, respectively, and
$\mathbf{F}=\mathbf{I}+\mathbf{J}$ with the projection $M_{F}$.

The magnetic dipole hyperfine constant ($A$) comes from the magnetic
dipole hyperfine operator $\mathcal{T}_{q}^{(1)}$ which is a tensor
of rank $1$. For an eigen state $\left|IJ\right\rangle $ of the
Dirac-Coulomb Hamiltonian, $A$ is defined as

\begin{equation}
A=\mu_{N}\left(\frac{\mu_{I}}{I}\right)\frac{\left\langle J\right\Vert \mathcal{T}^{(1)}\left\Vert J\right\rangle }{\sqrt{J(J+1)(2J+1)}},\label{mag-dip}\end{equation}
 where $\mu_{I}$ is the nuclear dipole moment defined in units of
Bohr magneton $\mu_{N}$. The magnetic dipole hyperfine operator $\mathcal{T}_{q}^{(1)}$
can be expressed in terms of single particle rank $1$ tensor operators
and is given by the first order term of Eq. (\ref{hyp-energy})

\begin{equation}
\mathcal{T}_{q}^{(1)}=\sum_{q}t_{q}^{(1)}=\sum_{j}-ie\sqrt{\frac{8\pi}{3}}r_{j}^{-2}\overrightarrow{\alpha_{j}}\cdot\mathbf{Y}_{1q}^{(0)}(\widehat{r_{j}})\,.\label{T1}\end{equation}
 Here $\overrightarrow{\alpha}$ is the Dirac matrix and $\mathbf{Y}_{kq}^{\lambda}$
is the vector spherical harmonics. The index $j$ refers to the $j$th
electron of the atom with $r_{j}$ as its radial distance and $e$
as the magnitude of the electronic charge.

The corresponding shift in the energy level is known as magnetic dipole
hyperfine shift and is expressed as

\begin{equation}
W_{M1}=A\frac{F(F+1)-I(I+1)-J(J+1)}{2}.\label{hyp-shift}\end{equation}

\section{\label{results}Results and discussions}

The transition which can serve as a new frequency standard is the
forbidden electric quadrupole ($E2$) transition $5d^{9}6s^{2}\,^{2}D_{5/2}\longrightarrow5d^{10}6s^{1}\,^{2}S_{1/2}$
in $\,^{199}Hg^{+}$. The possible shift which are crucial for accurate
determination of the desired transition frequency are mainly of two
kinds : electric quadrupole shifts of the $5d^{9}6s^{2}\,^{2}D_{5/2}$
state and the magnetic dipole hyperfine shifts of both the $5d^{9}6s^{2}\,^{2}D_{5/2}$
(upper) and the $5d^{10}6s^{2}\,^{1}S_{0}$ (lower) states. Departure
of the charge distribution from spherical nature to non-spherical
in the $5d^{9}6s^{2}\,^{2}D_{5/2}$ state will give rise to an electric
quadrupole moment which eventually will produce a shift in the energy
level in the presence of an external electric field gradient. On the
other hand, the non-zero nuclear spin of the $\,^{199}Hg^{+}$ which
causes the non-zero nuclear dipole moment (multipole moment of the
first kind) will give rise to the magnetic dipole hyperfine effect
in the presence of the electron multipole moment at the site of the
nucleus caused by the electron spin. This magnetic dipole hyperfine
effect will produce a shift in the energy levels for both (upper and
lower) states and are directly related to the magnetic dipole hyperfine
constant ($A$) which is given by Eq. (\ref{hyp-shift}).

In this paper we have calculated electric quadrupole moment ($\Theta$)
of the $5d^{9}6s^{2}\,^{2}D_{5/2}$ state and magnetic dipole hyperfine
constant ($A$) of the $5d^{9}6s^{2}\,^{2}D_{5/2}$, $5d^{10}6s^{2}\,^{1}S_{0}$
and $5d^{9}6s^{2}\,^{2}D_{3/2}$ states of $\,^{199}Hg^{+}$ using
relativistic Fock-space unitary coupled cluster theory. FSUCC theory,
is much more rigorous than its ordinary counterpart (Fock-space coupled
cluster theory, namely FSCC) and other atomic many-body theories like
configuration interaction , many-body perturbation theory etc. The
accuracy of our this calculation establishes the necessity of applying
a theory of this kind to calculate properties for complicated $D$-states.
In our calculation we have considered the ground state ($5d^{10}6s^{2}\,^{1}S_{0}$)
of $\,^{199}Hg$ as the Dirac-Fock (DF) reference state. We then apply
the closed shell unitary cluster operator to correlate the ground
state which is followed by a core ionization calculation to produce
the open shell states ($5d^{9}6s^{2}\,^{2}D_{3/2}$ and $5d^{9}6s^{2}\,^{2}D_{5/2}$)
of $\,^{199}Hg^{+}$. The basis functions are constructed by using
a large finite basis set expansion of Gaussian type orbitals \cite{rajat-gauss}
with $s,p,d,f$ and $g$ functions ($34s32p30d20f15g$). The nucleus
is assumed to have a finite structure (Fermi type). This closed shell
correlation calculation is followed by OSCC-IP \cite{csur-coreip}
calculations to obtain the $5d^{9}6s^{2}\,^{2}D_{5/2}$, $5d^{9}6s^{2}\,^{2}D_{3/2}$
states and an OSCC-EA \cite{mukherjee-oscc} calculation for obtaining
the $5d^{10}6s^{2}\,^{1}S_{0}$ state of $\,^{199}Hg^{+}.$ Excitations
from all the core orbitals have been considered to do a complete correlation
treatment. We have also studied the effects of higher angular momentum
states and found that to be negligible. Therefore we have omitted
the higher order symmetries to generate the basis functions. In an
earlier paper by one of the authors \cite{rkc-theocem}, the FSCC
method has been employed to estimate these quantities. In principal,
FSUCC is more rigorous than FSCC and contains more higher order effects
in the same level of approximation because of the unitary structure
of the closed shell correlation operator.

\begin{table}

\caption{\label{quad-moment}Electric quadrupole moment (${\bf \Theta}$ in
$ea_{0}^{2}$) of the $5d^{9}6s^{2}\,^{2}D_{5/2}$ state of $\,^{199}Hg^{+}$.
Entry within the parenthesis correspond to dressed one-body contribution.
FSUCC stands for the present calculation. MCHF and MCDF correspond
to multi-configuration Hartree-Fock and multi-configuration Dirac-Fock
(relativistic MCHF) respectively and `Expt.' is the experimental value.}

~

\begin{centering}
\begin{tabular}{ccccc}
\hline 
FSUCC&
HF &
MCHF &
MCDF&
Expt. \tabularnewline
\hline
\hline 
-0.517&
-0.664 \cite{Q-HF-itano}&
-0.544 \cite{itano-nist,Q-exp}&
-0.56374 \cite{itano-pra-06}&
-0.510 (18) \cite{Q-exp}\tabularnewline
(-0.739)&
&
&
&
\tabularnewline
&
&
&
&
\tabularnewline
\hline
\end{tabular}
\par\end{centering}
\end{table}

This particular calculation with $\,^{199}Hg^{+}$ is much more challenging
than the treatment of alkal-metal atoms and alkali like ions. The
$d$-shell vacancies in the excited states of $\,^{199}Hg^{+}$ introduce
additional complexities in the determination of atomic states and
related properties. The core ionization technique has been used in
connection with the FSUCC method for the first time to study the one-electron
properties of ions of this kind.

The earlier calculation \cite{rkc-theocem} by one of the authors
using FSCC has estimated the electric quadrupole moment (${\bf \Theta}$
in $ea_{0}^{2}$) of $5d^{9}6s^{2}\,^{2}D_{5/2}$ state of $\,^{199}Hg^{+}$
as $0.527$ $ea_{0}^{2}$, which was off by $0.017\, ea_{0}^{2}$
from the experiment (neglecting the experimental uncertainty). The
present estimate of ${\bf \Theta}$, on the other hand, is off by
$0.007\, ea_{0}^{2}$ from the central experimental value. We observe
that both the results (FSCC and FSUCC) are within the experimental
uncertainty ($\sim3.5\%$). \textbf{}However, it can be concluded
that the accuracy obtained in the present case using FSUCC will help
to reduce the experimental uncertainty. This can be understood by
observing the fact that at a given level of approximation (for this
case singles and doubles, namely SD) FSUCC theory contains much more
correlation and higher order excitation effects compared to FSCC.
Therefore applying an improved theory of this kind will give some
valuable inputs to the frequency standard measurements using $\,^{199}Hg^{+}$.

In a recent calculation Itano \cite{itano-pra-06} used the multi-configuration
Dirac-Fock (MCDF) method to estimate ${\bf \Theta}$ (in $ea_{0}^{2}$)
of the same state which gave the value to be $0.56374$ (this disagrees
with the experimental value by $\sim10.5\%$). The numbers of the
electric quadrupole moments ${\bf \Theta}$ (in $ea_{0}^{2}$) of
the $5d^{9}6s^{2}\,^{2}D_{5/2}$ and $5d^{9}6s^{2}\,^{2}D_{3/2}$
states of $\,^{199}Hg^{+}$are given in table \ref{quad-moment}.

We have also used FSUCC to determine the magnetic dipole hyperfine
constants ($A$) of the $5d^{9}6s^{2}\,^{2}D_{5/2}$, $5d^{9}6s^{2}\,^{2}D_{53/2}$
and $5d^{10}6s^{1}\,^{2}S_{1/2}$ states of $\,^{199}Hg^{+}$. Precise
calculations of $A$ values are not only theoretical checks of the
experimental determination but also provide information about the
accuracies of the atomic wavefunctions which are used in this calculation.
Table \ref{Hyp-A} contains the tabulated values of the $A$ using
different methods. The previous result obtained by FSCC \cite{rkc-theocem}
by one of the authors, reveals that the FSUCC is able to produce a
significant improvement over all of the values calculated here and
the theoretical accuracies have been reduced by significant amounts.
Experimental value of $A$ for the $5d^{9}6s^{2}\,^{2}D_{3/2}$ state
is not available. The accuracies of the entire calculation ensure
that the prediction of the $A$ value of the $5d^{9}6s^{2}\,^{2}D_{3/2}$
state will be able to lead to a precise measurement of the same.

From the given results in table (\ref{quad-moment}) and (\ref{Hyp-A})
the effects of electron correlation in determining electric quadrupole
moment and hyperfine constants can be estimated easily. The numbers
in the parentheses correspond to the effects of the dressed one-body
operator and the difference between the two values turns out to be
the contribution from electron correlation (many-body) effects. For
the electric quadrupole moment the correlation effects turns out to
be $\sim43\%$ and for the hyperfine constants the correlation contributions
range from $\sim2\%$ to $\sim19\%$. These effects clearly establish
the power of the FSUCC method to very accurately determine the atomic
properties, such as electric quadrupole moment and hyperfine constants.

\begin{table}

\caption{\label{Hyp-A}Magnetic dipole hyperfine constant ($A$ in MHz) of
different states of $\,^{199}Hg^{+}$. Entry within the parenthesis
correspond to dressed one body contribution. FSUCC stands for the
present calculation. The $\mu_{N}\left(\frac{\mu_{I}}{I}\right)$
values used in the calculation is taken from Ref. \cite{webelements}.}

~

\begin{centering}
\begin{tabular}{ccccc}
\hline 
State&
FSUCC&
FSCC \cite{rkc-theocem}&
MCDF \cite{itano-pra-06}&
Expt. \cite{Q-exp}\tabularnewline
\hline
\hline 
$5d^{9}6s^{2}\,^{2}D_{5/2}$&
995&
972&
963.5 &
986.19 (4) \tabularnewline
&
(865)&
&
&
\tabularnewline
$5d^{9}6s^{2}\,^{2}D_{3/2}$&
2780&
2713&
2478.3 &
\tabularnewline
&
(2734)&
&
&
\tabularnewline
$5d^{10}6s^{1}\,^{2}S_{1/2}$ &
40487&
40440&
&
40507\tabularnewline
&
(32963)&
&
&
\tabularnewline
\hline
\end{tabular}
\par\end{centering}
\end{table}

\section{\label{concl}Conclusion}

In summary, we have used the FSUCC method to determine the electric
quadrupole moment of the $5d^{9}6s^{2}\,^{2}D_{5/2}$ state and magnetic
dipole hyperfine constants of the $5d^{9}6s^{2}\,^{2}D_{5/2}$, $5d^{9}6s^{2}\,^{2}D_{3/2}$
and $5d^{10}6s^{1}\,^{2}S_{1/2}$ states of $\,^{199}Hg^{+}$. The
calculation involving $\,^{199}Hg^{+}$ is very complex and challenging
because the excited states which are of interests involve open $d$-
shells. The calculated values of the hyperfine constants can be considered
as benchmarking of the determination of atomic states using FSUCC
theory. Therefore, the accuracy obtained for the electric quadrupole
moments using this approach can help us to determine the uncertainty
which is very important to determine for the frequency standard studies.
To our knowledge this is the most accurate theoretical determination
of these quantities to date and are very important in the context
of producing a next generation frequency standard. FSUCC along with
the core ionization method has been applied to determine any atomic
properties of these kinds. This calculation will serve not only as
a supplement to the ongoing experiments but also to establish this
theory in determining different problems with atoms and ions.

\begin{verse}
\textbf{Acknowledgments} : This work is partially supported by the
National Science Foundation and the Ohio State University (CS). RKC
acknowledges the Department of Science and Technology, India (Grant
\# SR/S1/PC-32/2005). We sincerely acknowledge the constructive comments
made by the anonymous referee. 
\end{verse}

\end{document}